# Clinical Pathways Matter for Multimodal Deep Learning in Early Alzheimer's Disease Detection

Yao Lu[1,2], Solveig Kristina Hammonds[1,3,4] and Alvaro Fernandez-Quilez[1,*], for the Alzheimer's Disease Neuroimaging Initiative

[1]Department of Computer Science and Electrical Engineering, University of Stavanger, Kjell Arholms Hus 41, 4021, Stavanger, Norway.

[2]Department of Computer Science, University of Putra Malaysia, Jalan Universiti 1, 43400, Selangor, Malaysia.

[3]SMIL, Department of Radiology, Stavanger University Hospital, Gerd-Ragna Bloch Thorsens Gate 8, 4011, Stavanger, Norway.

[4]Centre for Age-Related Medicine (SESAM), Stavanger University Hospital, Gerd-Ragna Bloch Thorsens gate 8, Stavanger, 4011, Norway

[*]Corresponding author: Alvaro Fernandez-Quilez, Department of Computer Science and Electrical Engineering, University of Stavanger, Kjell Arholms Hus 41, 4021, Stavanger, Norway. E-mail: alvaro.f.quilez@uis.no

**Abstract**

Identifying individuals at risk of Alzheimer's disease (AD), particularly in the preclinical and early stages, remains challenging. Although deep learning approaches based on structural MRI show promise as a non-invasive biomarker, existing multimodal models require task-specific training and depend on biomarkers that are not routinely available in clinical practice. Here, we propose a zero-shot multimodal framework based on SigLIP that combines structural MRI embeddings with text embeddings of routinely collected clinical variables for early AD risk stratification in individuals at preclinical or mild cognitive impairment (MCI) stages. We evaluated the approach in 416 individuals from the ADNI cohort (age: 72.73 ± 6.7). SigLIP was used without fine-tuning to extract MRI and clinical text embeddings, which were combined into multimodal representations for individual-level AD risk prediction within 4 years. We further compared the model performance in a single-visit and two-visit settings to assess the value of longitudinal information and framework scalability. In the single-visit setting, combining MRI embeddings with MMSE, age, and sex achieved an AUC of 0.91 ± 0.02, outperforming both a CSF Aβ42-based model (AUC 0.73 ± 0.08) and an MMSE-based model (AUC 0.85 ± 0.22). In the two-visit setting, performance was maintained or improved, supporting the scalability of the approach to longitudinal data. These findings suggest that zero-shot multimodal fusion of structural MRI and routinely collected clinical variables may provide a practical and scalable strategy for early AD risk stratification without task-specific retraining.

## 1. BACKGROUND

Alzheimer’s disease (AD) is a progressive neurodegenerative disorder that evolves over years, typically transitioning from a preclinical phase to mild cognitive impairment (MCI) and dementia [1]. In developed countries, AD affects approximately 1 in 10 older adults aged 65 and above, and over a third of those aged 85 and older [2]. Although early detection could support timely intervention and improve trial stratification, it remains challenging because early markers are subtle and individual clinical trajectories are heterogeneous [3].

In clinical practice, assessment is anchored in patient history, cognitive testing and functional evaluation [4]. In that regard, cognitive test instruments such as the Clinical Dementia Rating (CDR) and the Mini-Mental State Examination (MMSE) can support staging and trial enrolment but their sensitivity is limited when impairment is mild [5]. Structural MRI is often acquired as part of the routine evaluation of cognitive symptoms, primarily to exclude alternative causes of impairment such as vascular lesions [6]. Although MRI can also reveal patterns of neurodegeneration consistent with AD, these changes may be subtle early on and their interpretation depends on expertise and acquisition factors [6,7].

Deep learning (DL) has shown promise for extracting imaging features relevant to AD from MRI [3,8,9], potentially capturing complex patterns beyond predefined regions or voxel-wise statistics [10–12]. Existing works leverage 3D convolutional networks or more recent transformer-based architectures to T1-weighted MRI for AD-versus-control classification and for predicting MCI-to-AD conversion [13,14]. However, performance is typically highest in established dementia and degrades in prodromal cohorts, where disease-related changes are subtle and heterogeneous across individuals [15]. This has motivated multimodal approaches that integrate imaging with demographic and clinical measures to approximate clinical reasoning more closely [16,17].

Despite their promise, multimodal models are highly sensitive to clinical-variable choice and collection context [18]. In practice, clinical measurements are not universally available or synchronously acquired. The collection timing can vary relative to imaging, missingness is systematic, and cohorts reflect referral patterns and diagnostic escalation [19]. In that regard, a substantial fraction of multimodal work incorporates specialized or invasive biomarkers such as PET imaging or cerebrospinal fluid (CSF) [20]. These may be costly, invasive, or not routinely available in standard care [21]. Not accounting for those factors can yield optimistic performance estimates while limiting clinical use and adoption [8].

Here, we propose a multimodal DL framework that is grounded in the clinical pathway of patients with clinically suspected dementia and integrates routinely collected point-of-care clinical information with structural MRI, explicitly accommodating asynchronous and incomplete data. We encode structured clinical variables as compact text descriptors and use Sigmoid Loss for Language–Image Pre-training (SigLIP) to align the clinical text descriptors with MRI-derived embeddings [22]. We evaluate our approach against unimodal baselines and an invasive reference biomarker (CSF Aβ42) to quantify how much diagnostic signal can be recovered from routine clinical variables when clinical context is respected. We hypothesize that pathway-aware multimodal fusion will improve early AD detection relative to MRI-only and routine-clinical-only models, achieving performance comparable to approaches using CSF Aβ42.

## 2. METHODS

### 2.1 Participants and study design

The data used in this study was sourced from Alzheimer's Disease Neuroimaging Initiative (ADNI; adni.loni.usc.edu), a large, ongoing longitudinal study launched in 2003 to understand the progression of AD from normal aging to MCI and early AD [23]. Participants were recruited in-person from clinical centers across North America. The present study utilized data from ADNI-1, ADNI-GO, ADNI-2, and ADNI-3.

The ADNI database provides a comprehensive collection of multimodal data, including extensive imaging, CSF and blood biomarkers, cognitive test scores, genetic profiles, and detailed clinical and demographic information. In our study, these resources were utilized to evaluate the predictive value of clinical and imaging features for AD progression. The analyzed cohort comprised participants categorized as cognitively normal (CN), MCI, or mild AD, in line with ADNI diagnostic protocols.

To ensure consistency in imaging quality and avoid confounding by unwanted clinical factors, we applied the following inclusion criteria: (1) individuals who had a baseline 1.5T or 3.0T T1-weighted MP-RAGE MRI scan and no dementia-unrelated neurological or psychiatric disorders; (2) diagnosed as either MCI or CN at the first visit; (3) at least two available visits; (4) clinical diagnosis, and age available at the first visit; (5) baseline MRI scans of diagnostic quality as defined in [23]. After applying these criteria, eligible participants were divided into two groups based on clinical progression over up to 4 years: converters, who progressed from CN or MCI at baseline to a diagnosis of probable AD at any follow-up visit, and non-converters, who retained their baseline diagnosis throughout the follow-up period. A total of 416 individuals met all inclusion criteria. The full selection process and resulting cohort is illustrated in Figure 1.

**2.2 MRI and pre-processing**

T1-weighted MRI scans were collected from participants scanned on either 1.5T or 3.0T scanner systems. The images were acquired across multiple scanner vendors between September 2002 and April 2022 using commercial scanners manufactured by Siemens Healthineers (Erlangen, Germany), Philips Medical Systems (Eindhoven, The Netherlands), and GE Healthcare (Chicago, Illinois, USA). ADNI applies standardized MRI protocols depending on field strengths, head coil and scanner type, with a nominal FOV of approx. 240 x 240 $mm^2$, number of slices in the range of 160-208, and a voxel size of approx. 1 x 1 x 1 $mm^3$ [24]

To reduce inter-subject and inter-scanner variability, all images were preprocessed using a Python-based pipeline adapted from the open-source ADNI preprocessing toolkit (adni_preprocessing). In our study, the order of preprocessing steps was adjusted to better accommodate the characteristics of our dataset. MRI preprocessing was performed with the SimpleITK library (version 2.4.1) [25]. The preprocessing steps included: (1) N4 bias field correction [26] to address low-frequency intensity non-uniformities (2) skull stripping to remove non-brain tissue and facilitate inter-subject brain alignment (3) registration to the MNI152 T1-weighted template [27] using a mutual information–based cost function and B-spline interpolation to preserve anatomical structures (4) cropping from the original field of view 208×240×256 mm³ to a standardized volume of 182×218×182 mm³, ensuring spatial consistency across participants. All images after registration contained at least 182 slices in the z-direction, ensuring that no anatomical information was lost. All pre-processing steps and resulting images are presented in Figure 2a.

**Transformation from 3D MRI to 2D representations**

The original input data consist of preprocessed 3D T1-weighted MRI volumes. Since the SigLIP vision encoder operates on 2D images, each 3D volume was transformed into a set of 2D axial slices and each slice was used independently as input to the SigLIP model. This allowed the model to extract visual features from each slice of the 3D volume.

The process of transformation is as shown in Figure 2B. For each subject, axial slices corresponding to the central portion of the volume, approximately the middle 1/3 to 2/3 of slices were selected [28]. This range was chosen to capture anatomically informative regions, including the hippocampus and medial temporal lobe, which are known to exhibit early structural changes in AD [2,29,30]. These 2D slice representations were then used as inputs to the SigLIP pipeline for visual feature extraction.

**2.3 Multimodal deep learning-based analysis**

We developed a multimodal DL model that utilizes pretrained SigLIP [22] encoders used solely for feature extraction without further training to obtain visual and textual embeddings from 2D T1-weighted MRI slices and corresponding patient demographic information and neuropsychological tests. For each subject, all visual features extracted from the MRI slices corresponding to the same MRI volume were aggregated to form a single, subject-level imaging representation (Figure 2B). Image features extracted from individual slices were initially stacked along the slice dimension to form a feature matrix, where each row represents the feature vector of a single slice. This matrix was then subjected to an averaging process, where the features across all slices of a given subject were averaged along the slice axis. The result of this is a single image feature vector for each subject that integrates the visual information from all the relevant slices (Figure 2B). This aggregation allows predictions to be made at the subject level, consolidating the anatomical information of a subject across multiple slices into a unified representation.

The structured patient history and clinical data (age, sex, MMSE and CDR score) in the ADNI database were tokenized using SigLIP's built-in tokenizer [22]. After tokenization, the resulting text was passed into SigLIP's text encoder, which produced a 512-dimensional embedding for each sample. Each individual slice within this range was treated as a separate input sample and paired with the corresponding subject's clinical description for multimodal classification. The image and text embeddings were concatenated to form a 1024-dimensional multimodal feature vector. The combined features were subsequently used to train a logistic regression model [31] to predict the probability of AD progression within 4 years. (Figure 2B). Logistic regression was selected as a clinically well-established and interpretable model, providing a standardized and simple predictive framework that allows the contribution of SigLIP-derived embeddings to be evaluated without confounding effects introduced by more complex classifiers. This design ensures that observed performance differences primarily reflect the effectiveness of the extracted multimodal representations rather than variations in model complexity.

A total of 2718 MRI scan sessions from the 416 participants were included. To examine the effect of additional longitudinal data, we controlled the number of visits per subject. Specifically, two settings were compared: one using only the baseline visit (1-visit setting), and another including up to 2 visits (2-visit setting). This allowed us to assess the influence and scalability of SigLiP when using longitudinal data in the 4-year AD progression risk estimation.

Training and test sets were split at the subject level to avoid cross-contamination, with 80% of the subjects for training (332 subjects) and 20% for testing (84 subjects). All slices belonging to a single subject and all participants' visits were contained exclusively in either training or testing set, ensuring that no data from the same individual appeared in both training and testing to prevent possible data leakages.

All experiments were implemented using Python (version 3.10), PyTorch (version 2.5.1), and the Hugging Face transformers library (version 4.52.1). The logistic regression classifier implemented with Scikit-learn (version 1.0.2), was trained using the Limited-memory Broyden–Fletcher–Goldfarb–Shanno (LBFGS) optimizer with a maximum of 500 iterations (max_iter=500) to ensure convergence. All training and inference procedures were conducted on a Windows 11 workstation with an Intel Core i5-1340P CPU and 16GB RAM.

## 2.4 Evaluation

### AD progression risk

To evaluate the predictive performance of the models across different experimental conditions, we adopted balanced accuracy (BACC), area under the ROC curve (AUC), and F1 score. These metrics offer complementary perspectives on model behavior and support a more robust assessment when the available dataset is limited [32-34]. In addition, confusion matrices are also presented to study in detail sensitivity and specificity trade-offs at a given clinically-relevant false positive rate (FPR) [34].

### Additive effect of available amount of point-of-care information

We conducted a series of experiments based on image and text features extracted by the SigLIP model to systematically evaluate the impact of different point-of-care and commonly available clinical and demographic patient information. Specifically, we considered the following combinations of text features with SigLIP: (a) CDR and age, (b) CDR, age and sex (c) MMSE and age, and (d) MMSE, age and sex. In all cases, combinations were tested jointly with MRI. These combinations were selected to simulate a range of clinically meaningful scenarios with different amounts of available clinical information. To further reflect realistic clinical settings, we additionally evaluated the same combinations of clinical and demographic features within the same SigLIP-based framework but without MRI input (Appendix Table S3 and Table S4) in order to quantify the incremental effect of MRI. This experimental setup allows us to address how clinically asynchronous and different combinations of structured information affect 4-year risk estimation performance of AD.

**Text-encoding strategies**

For each feature combination, we evaluated three encoding strategies to convert structured variables into text inputs: (a) textual descriptions only, (b) numeric values only, and (c) a combined format including both numeric values and textual descriptions (Table S1). This approach aimed to have a systematic understanding of how different encoding methods influence the model's ability to capture meaningful patterns from clinical and demographic features encoded as text. Specifically, we assessed the individual contributions of each strategy to predictive performance and examined whether the text encoder could effectively utilize both contextual information and numerical patient data. As a representative case to depict the encoding process and experimentation with the variables, consider the variable CDR and the effect of different encoding strategies: (a) CDR value presented as plain numeric string "0.5", directly presenting the severity rating without contextual information; (b) a semantic-only representation was used "Clinical Dementia Rating:", which reflects a typical clinical phrase but omits the numeric value; (c) a hybrid format was adopted that combines both the numerical score and descriptive text, "Clinical Dementia Rating: 0.5", to better resemble real-world clinical notes and patient history. For each feature combination, we maintained the same dataset splits and image features across different encoding strategies, ensuring that only the textual representation varied.

**Baselines**

To establish a baseline performance, we implemented logistic regression on both individual and combined clinical variables in the form of a numerical input and with no further encoding. This included: (a) CDR, (b) MMSE, (c) CSF Aβ42, (d) CSF Aβ42 with CDR and age, (e) CSF Aβ42 with CDR, age and sex, (f) CSF Aβ42 with MMSE and age, and (g) CSF Aβ42 with MMSE, age and sex. These baseline models reflect the relative accessibility of clinical features in practice: CDR, MMSE, age, and sex can be easily obtained through routine assessments and provide a reference for evaluating the potential benefits of multimodal integration with SigLIP. We leverage CSF Aβ42 as our main clinical reference standard, as that is typically the standard for AD confirmation (Table 3) [4].

**2.5 Statistical testing**

Continuous variables that were normally distributed were summarized as mean ± standard deviation (SD). Categorical variables were reported as counts and percentages. Group comparisons were conducted using parametric or non-parametric tests as appropriate based on distributions.

To quantify performance variability, we used a stratified bootstrapping approach. For each model, we performed 100 bootstrap iterations with resampling stratified by AD progression. All performance metrics are calculated for each re-sampled sub-set, resulting in a distribution of 100 scores per metric. Based on these bootstrap distributions, performance is reported as mean ± standard deviation in the tables and for ROC curve comparisons.

Statistical comparisons between models were performed using two-sided paired t-tests on bootstrapped performance distributions, and P values are reported for each comparison. After accounting and correcting for multiple testing, $P < 0.01$ were considered statistically significant. All statistical analyses were conducted in Python 3.10.1 (https://www.python.org/downloads/) with SciPy (https://scipy.org/) library.

# 3. RESULTS

## 3.1 Sample description

The demographic and clinical characteristics of all participants, as well as those of converters and non-converters are presented in Table 1. Our study included 416 individuals, with an average age of 72.73±6.70 years. Female participants made up a slightly lower portion of the cohort (49.04%) compared to male participants (50.96%). At the baseline visit, the majority of subjects were categorized as CN (61.06%). Among the participants, 81 subjects were identified as converters, while the remaining  were non-converters. On average, converters presented lower MMSE (26.41 ± 2.18) and higher CDR (0.50 ± 0.16) values than non-converters.

Figure 3(a) depicts that multimodal combinations outperform unimodal approaches in the 1-visit setting, as further reflected in the confusion matrices (Figure 3(b)). For the multimodal model with SigLIP MRI embeddings and clinical features (CDR, age, sex), confusion matrices at FPR thresholds of 0.1 and 0.2 were analyzed. At FPR = 0.1, the model identified 33 non-AD converters and 6 AD converters, with 3 false positives and 10 false negatives, indicating high specificity but limited sensitivity. When the FPR threshold was relaxed to 0.2, false positives increased to 6, false negatives decreased to 6, and correctly predicted AD cases rose to 10, improving sensitivity with a moderate trade-off in specificity.

## 3.2 Added value of multi-modal data

To investigate the effectiveness of multimodal information, we first focused on the model performance under the 1-visit setting (Table 3). SigLIP MRI alone did not perform as well as Aβ42 alone, but when combined with clinical features, such as CDR and age, SigLIP MRI showed a substantial improvement in performance, highlighting the complementary value of integrating imaging-derived features with clinical data. Among the multimodal combinations, the best performance was achieved by combining SigLIP MRI with MMSE, Age and Sex (AUC=0.91 ± 0.02), which outperformed both Aβ42 alone (AUC = 0.73 ± 0.08) and SigLIP MRI alone (AUC = 0.63 ± 0.08).

### 3.3 Effect of clinical features encoding

We conducted experiments with three types of encoding. Each encoding was integrated with Siglip-derived image features under both 1 visit and 2 visits . In the 1-visit condition (Table 2), numeric-only encoding performed best for SigLIP MRI with Age (AUC = 0.72 ± 0.18), while full-text encoding yielded the highest performance for SigLIP MRI with CDR and SigLIP MRI with CDR and age. In the 2-visit setting (Appendix Table S1), full-text encoding outperformed all other methods

### 3.4 Effect of additional visits

As illustrated in Figure 4, most models showed improved performance in the 2-visit setting compared to the 1-visit setting. The model of SigLIP MRI with CDR, age, and sex (AUC=0.92 ± 0.03) achieved the highest performance in the 2-visit condition (Appendix Table S2). However, the performance of SigLIP MRI with MMSE and age, and SigLIP MRI with MMSE, age and sex was slightly lower in the 2-visit setting compared to the 1-visit condition.

## 4. DISCUSSION

In this retrospective study, we systematically evaluated a series of SigLIP-based unimodal and multimodal classification models for predicting the 4-year risk of AD progression, highlighting the feasibility of integrating multimodal features through SigLIP in a manner that accounts for clinical availability. Our results demonstrate the complementary predictive value of combining MRI-derived features with clinical indicators across distinct patient pathways. To our knowledge, this is the first study to apply SigLIP to 4-year AD risk prediction leveraging a zero-shot pretrained vision-language model to derive semantically rich image representations without task-specific image-model training [35-38]. In contrast to prior work that has predominantly relied on invasive or costly biomarkers [39], such as cerebrospinal fluid amyloid and tau measurements or FDG–PET imaging [40], our study demonstrates the feasibility of multimodal integration through SigLIP under clinically realistic constraints of data availability. The results show that combining MRI-derived representations with clinical and demographic variables yields complementary predictive value, supporting a practical framework for non-invasive multimodal risk stratification across heterogeneous patient pathways.

Beyond the use of SigLIP itself, our study also contributes methodologically by examining how clinical information is represented and how limited longitudinal information can be incorporated. In most settings, full-text encoding performed better than representations based only on textual descriptors or numerical values, suggesting that preserving both qualitative and quantitative aspects of patient data may support more informative patient-level modelling [41]. Many previous studies reduce clinical variables to structured numerical inputs [15, 42], which may fail to capture information contained in richer clinical descriptions.Unlike many prior studies that relied on single-visit inputs [43,44], we also compared the scalability of our method by comparing one-visit and two-visit settings to assess whether incorporating an additional visit could enrich the available patient information. These results suggest that adding a second visit can provide complementary clinical context and a more comprehensive representation of disease status.. Together, these analyses extend the study beyond simple model comparison and offer insight into how different encoding strategies and the inclusion of additional clinically available data may influence AD progression prediction.

Whilst our SigLIP-based approach offers powerful representational capacity, several limitations must be acknowledged. This study was restricted to one-visit and two-visit settings, and the modest sample size may constrain generalizability. Slice-to-subject aggregation may also obscure spatially localized disease signals. In addition, SigLIP encoders were used without AD-specific fine-tuning, and task-adapted optimization may further improve performance. Larger independent cohorts will be essential to validate the robustness and clinical utility of this framework.

## 5. CONCLUSION

Our study demonstrates that structural MRI features extracted using SigLIP when combined with routinely available clinical biomarkers can serve as a non-invasive, stable, and effective approach for predicting individual AD progression. By leveraging zero-shot image feature extraction from a pretrained model, the framework supports early risk estimation in both single-visit and longitudinal settings, offering a scalable and accessible tool for clinical monitoring and intervention planning.

## DATA AVAILABILITY STATEMENT

The data used in this study were obtained from the Alzheimer’s Disease Neuroimaging Initiative (ADNI) database (http://adni.loni.usc.edu/), which is publicly available to qualified researchers upon application.

## ACKNOWLEDGMENTS

We thank the Alzheimer’s Disease Neuroimaging Initiative (ADNI) for providing the data used in this study. A complete list of ADNI investigators can be found at: https://adni.loni.usc.edu/wp-content/uploads/how_to_apply/ADNI_Acknowledgement_List.pdf.

## AUTHOR CONTRIBUTIONS

Y.L. designed the study, performed the experiments, curated and analyzed the data, and wrote the initial draft of the manuscript. S.K.H. critically appraised the conceptualization and development of the manuscript, reviewed the manuscript in all stages, assisted in writing and approved the final manuscript. A.F.Q. supervised the project, provided conceptual guidance, guided the experimental work, and approved the final manuscript. All authors read and approved the final manuscript.

## DATA AVAILABILITY STATEMENT

The data used in this study were obtained from the Alzheimer's Disease Neuroimaging Initiative (ADNI) database (http://adni.loni.usc.edu/). Researchers can apply for access to the ADNI dataset via the official website.

## ADDITIONAL INFORMATION

## COMPETING INTEREST STATEMENT

The authors declare that they have no competing interests.

## ETHICS DECLARATION

All procedures performed in studies involving human participants were in accordance with the ethical standards of the institutional research committee and with the 1964 Helsinki declaration and its later amendments. Informed consent was obtained from all individual participants included in the ADNI dataset.

## CONSENT FOR PUBLICATION

Not applicable.

## FUNDING

Open access funding provided by University of Stavanger. Data collection and sharing for this project was funded by the Alzheimer's Disease Neuroimaging Initiative (ADNI) (National Institutes of Health Grant U01 AG024904) and DOD ADNI (Department of Defense award number W81XWH-12-2-0012). ADNI is funded by the National Institute on Aging, the National Institute of Biomedical Imaging and Bioengineering, and through generous contributions from the following: AbbVie, Alzheimer's Association; Alzheimer's Drug Discovery Foundation; Araclon Biotech; BioClinica, Inc.; Biogen; Bristol-Myers Squibb Company; CereSpir, Inc.; Cogstate; Eisai Inc.; Elan Pharmaceuticals, Inc.; Eli Lilly and Company; EuroImmun; F. Hoffmann-La Roche Ltd and its affiliated company Genentech, Inc.; Fujirebio; GE Healthcare; IXICO Ltd.; Janssen Alzheimer Immunotherapy Research & Development, LLC.; Johnson & Johnson Pharmaceutical Research & Development LLC.; Lumosity; Lundbeck; Merck & Co., Inc.; Meso Scale Diagnostics, LLC.; NeuroRx Research; Neurotrack Technologies; Novartis Pharmaceuticals Corporation; Pfizer Inc.; Piramal Imaging; Servier; Takeda Pharmaceutical Company; and Transition Therapeutics. The Canadian Institutes of Health Research is providing funds to support ADNI clinical sites in Canada. Private sector contributions are facilitated by the Foundation for the National Institutes of Health (www.fnih.org). The grantee organization is the Northern California Institute for Research and Education, and the study is coordinated by the Alzheimer's Therapeutic Research Institute at the University of Southern California. ADNI data are disseminated by the Laboratory for Neuro Imaging at the University of Southern California.

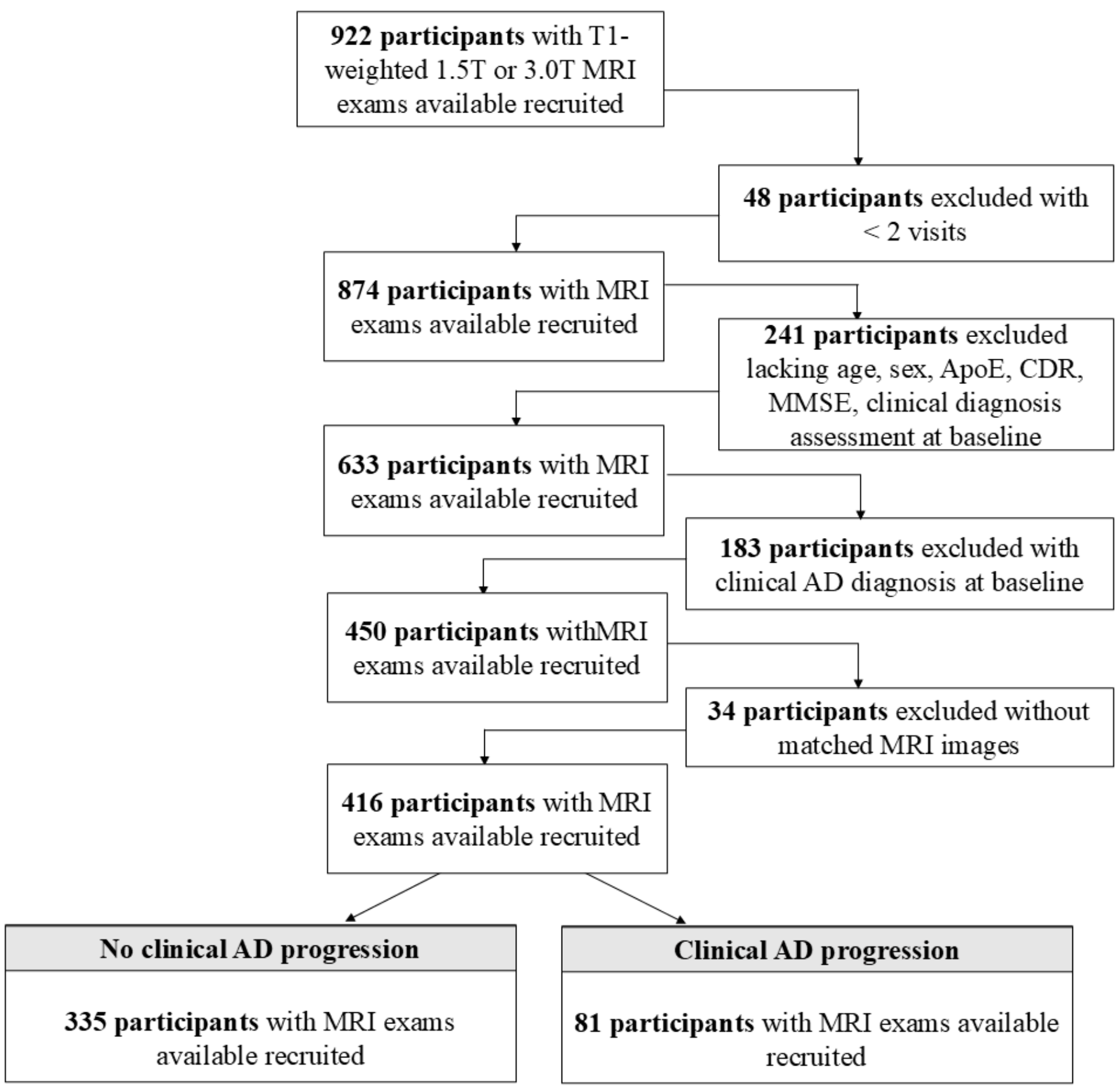

**Figure1**. Criteria for participant inclusion and the corresponding number of T1-weighted MRI exams. AD = Alzheimer's disease, CDR = Clinical Dementia Rating, MMSE = Mini-Mental State Examination, APOE = Apolipoprotein E.

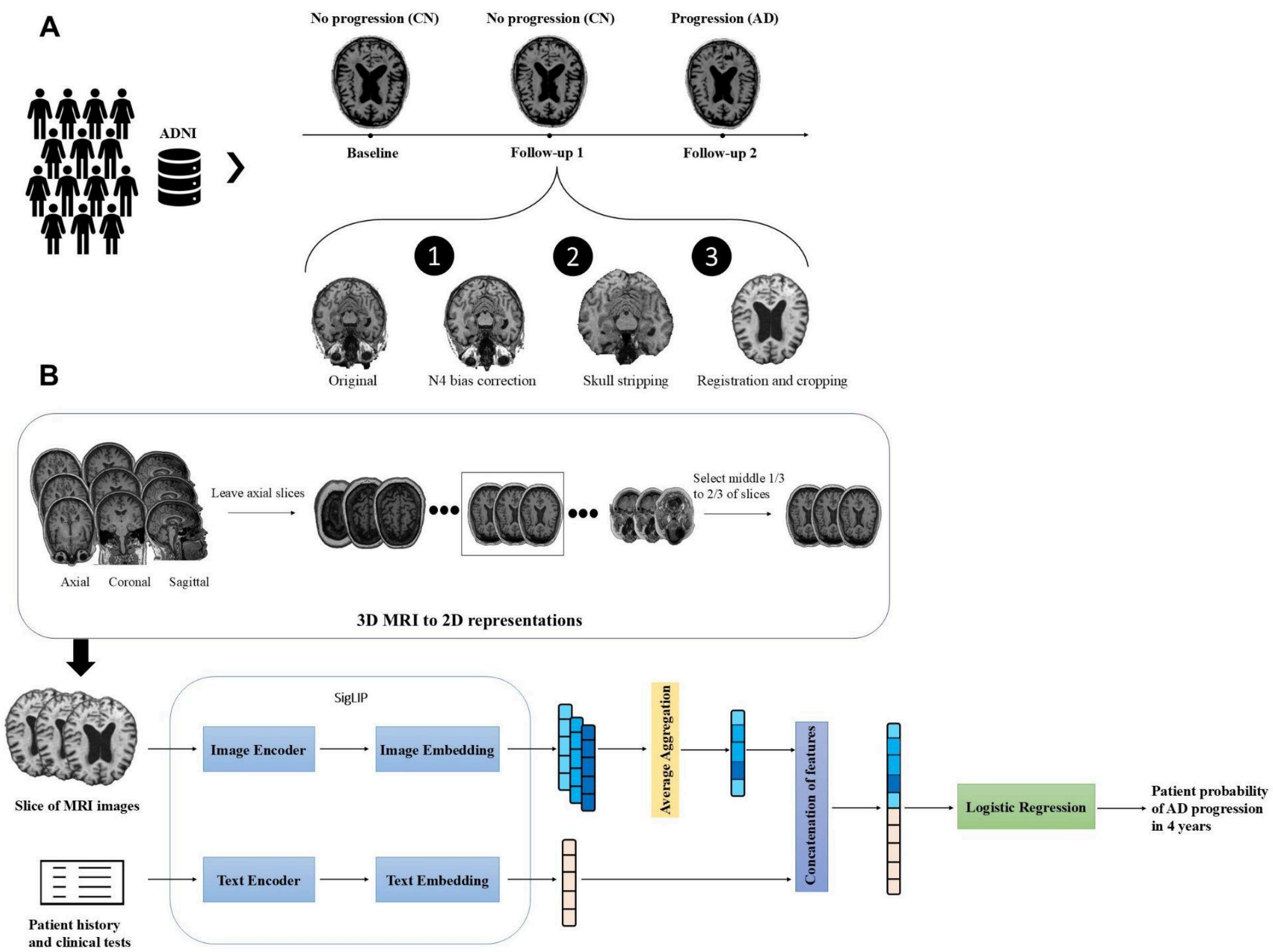

**Figure 2.** Method overview (A) ADNI dataset with longitudinal T1-weighted MRI data, including progression annotation and MRI pre-processing step (B) Multimodal classification pipeline using SigLIP to extract visual and textual embeddings from 2D T1-weighted MRI slices and clinical data, with concatenated features input into a logistic regression classifier for AD prediction. CN = Cognitively normal, AD = Alzheimer's disease.

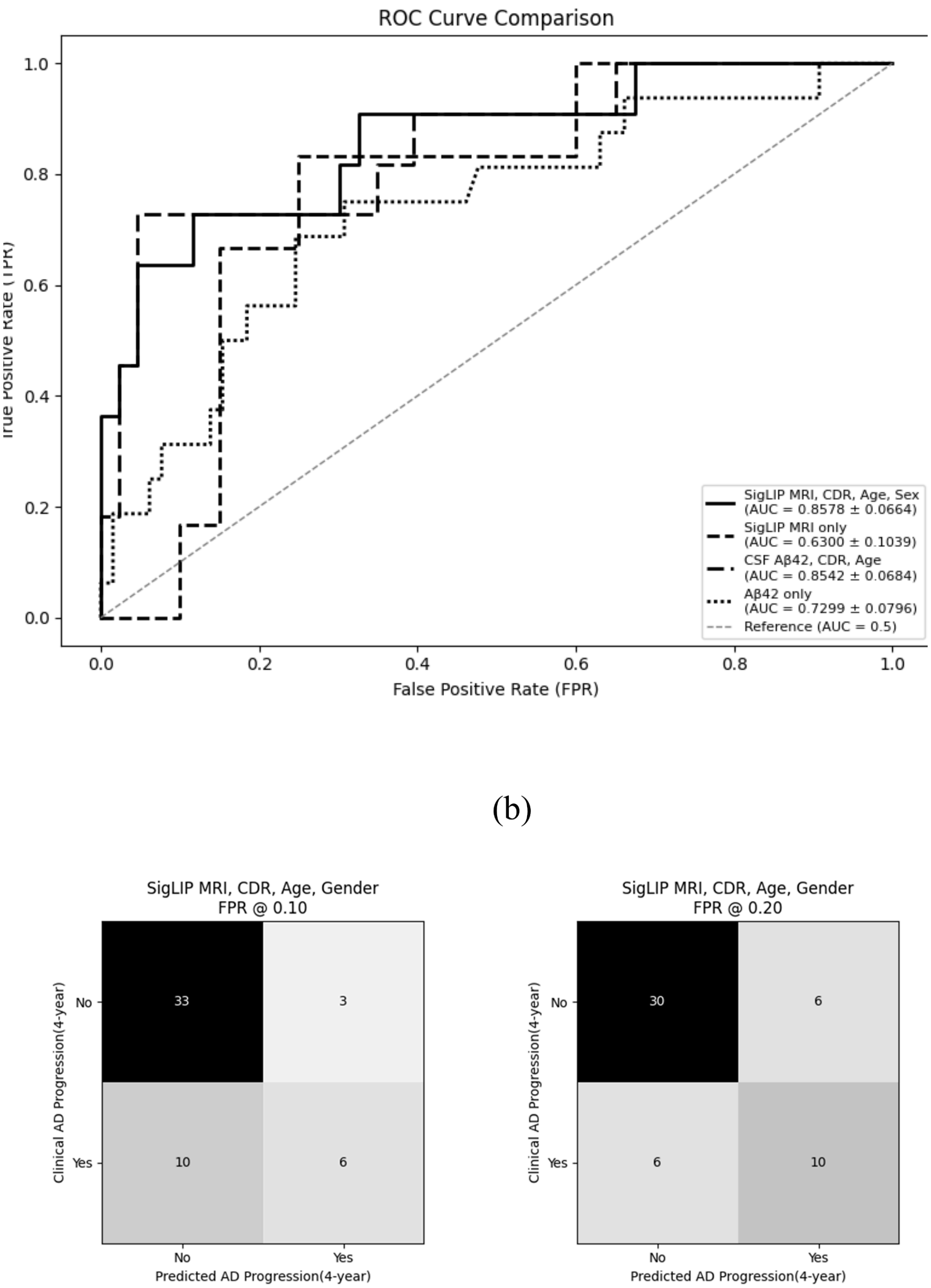

Figure 3 (a). Discriminative ability of four logistic regression models under the 1-visit setting, including unimodal models using only Aβ42 and only SigLIP MRI, and multimodal combinations of SigLIP MRI with CDR and age, and SigLIP with CDR, age and sex. (b)

Confusion matrices of the SigLIP-based multimodal model integrating image embeddings and clinical features (CDR, age, sex), under the 1-visit setting. The two matrices correspond to fixed FPR thresholds of 0.1 and 0.2 respectively, illustrating the trade-off between sensitivity and specificity at different decision thresholds. CDR = Clinical Dementia Rating. FPR= false positive rate.

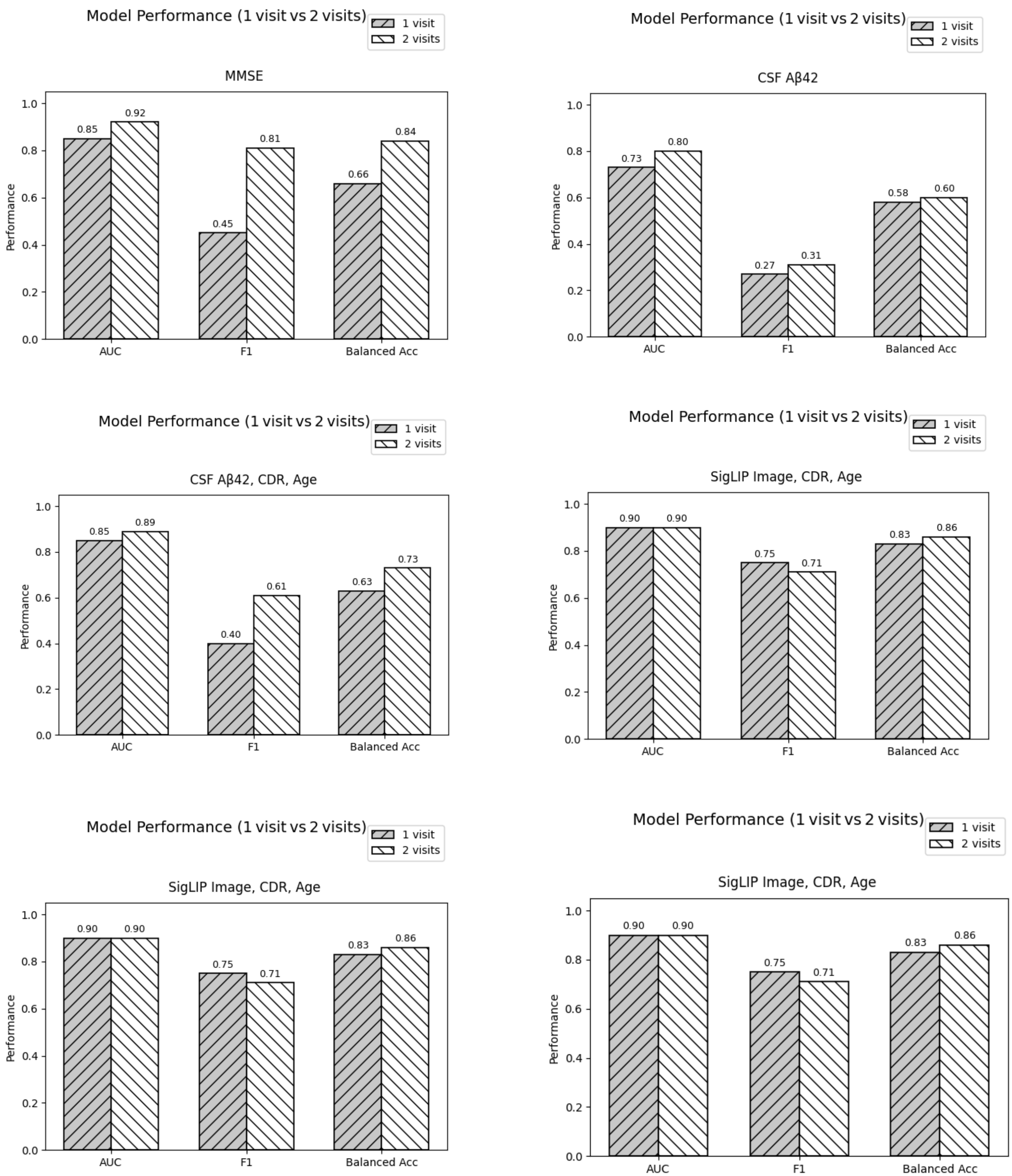

**Figure 4.** Comparative performance of models using one versus two visits. The models cover a range of unimodal and multimodal configurations, including clinical variables (CDR and MMSE ), combined clinical features (model with CSF Aβ42, CDR, age),and various multimodal combinations (SigLIP MRI with CDR, age, and sex). Performance is evaluated using AUC, F1 score, and balanced accuracy. CSF = Cerebrospinal fluid, CDR = Clinical Dementia Rating, MMSE = Mini-Mental State Examination.

| | All | Non-converters | Converters | Statistics |
|---|---|---|---|---|
| **Participants**<br>Total, N<br>Male:Female<br>Female % | <br>416<br>212:204<br>49.04 | <br>335<br>165:70<br>50.75 | <br>81<br>47:34<br>41.98 | <br><br><br>p=0.29 |
| **Baseline diagnosis**<br>CN:MCI<br>MCI, % | <br>254:162<br>38.94 | <br>250:85<br>25.37 | <br>4:77<br>95.06 | <br><br>p<0.01 |
| **Age** [years] | 72.76 ± 6.70 | 72.59 ± 6.76 | 73.29 ± 6.46 | p=0.39 |
| **Cognitive tests**<br>MMSE<br>CDR | <br>28.28 ± 1.93<br>0.22 ± 0.26 | <br>28.74 ± 1.55<br>0.16 ± 0.24 | <br>26.41 ± 2.18<br>0.50 ± 0.16 | <br>p<0.01<br>p<0.01 |
| **APOE genotype**<br>APOE4-,0,N<br>APOE4+ 1/2, N | <br>269<br>118/29 | <br>236<br>84/15 | <br>28<br>39/14 | <br><br>p<0.01 |

**Table 1.** Baseline demographic and clinical characteristics of all participants, along with data for converters and non-converters in the 4-year-follow up study. These characteristics were calculated separately for each group. This breakdown allows for a more detailed comparison of the baseline features between converters and non-converters. Non-converters: All records from individuals that are CN or MCI and never progress to AD. Converters: All records from individuals that are CN or MCI and progress to AD. APOE: Apolipoprotein E, CN: Cognitively normal, MCI: Mild Cognitive Impairment, CDR: Clinical Dementia Rating, MMSE: Mini-Mental State Examination. The p-values were computed for comparisons between non-converters and converters. p-values < 0.01 were considered statistically significant after multiple-testing correction.

| **Model** | **Textual descriptive only** | **Numeric only** | **Full text with numeric and textual descriptive** | **Statistics** |
|---|---|---|---|---|
| SigLIP MRI, Age<br>AUC | 0.55 ± 0.21 | 0.72 ± 0.18 | 0.70 ± 0.20 | p<0.01*<br>p<0.01** |
| SigLIP MRI, CDR<br>AUC | 0.63 ± 0.27 | 0.80 ± 0.08 | 0.84 ± 0.04 | p<0.01*<br>p<0.01** |
| SigLIP MRI CDR, Age<br>AUC | 0.65 ± 0.23 | 0.85 ± 0.07 | 0.90 ± 0.06 | p<0.01*<br>p<0.01** |

**Table 2.** AUC Performance of multimodal models using different text encodings with SigLIP MRI under the 1-visit setting. All experiments were conducted on the same dataset splits and MRI features to ensure fair comparison. Results are computed using 5-fold cross-validation, with bootstrapping (N = 100) and presented as mean ± standard deviation. [95% CI]. The probability threshold was optimized with respect to balanced accuracy for every case. CSF = Cerebrospinal fluid, CDR = Clinical Dementia Rating, MMSE = Mini-Mental State Examination. * is the p-value for textual descriptive only, calculated using full text with numeric and textual descriptive as reference, ** is the p-value for numeric only, calculated using full text with numeric and textual descriptive as reference, p-values<0.01 were considered statistically significant after correcting for multiple testing.

| Model | AUC | F1 | Balanced accuracy | Statistics |
|---|---|---|---|---|
| Baseline CDR | 0.84 ± 0.08 | 0.29 ± 0.26 | 0.59 ± 0.09 | p<0.01 |
| Baseline MMSE | 0.85 ± 0.22 | 0.45 ± 0.65 | 0.66 ± 0.29 | p<0.01 |
| **Invasive baseline** | | | | |
| CSF Aβ42 | 0.73 ± 0.08 | 0.27 ± 0.22 | 0.58 ± 0.08 | p<0.01 |
| CSF Aβ42, CDR, Age | 0.85 ± 0.07 | 0.40 ± 0.28 | 0.63 ± 0.12 | p<0.01 |
| CSF Aβ42, CDR, Age, Sex | 0.86 ± 0.11 | 0.53 ± 0.18 | 0.73 ± 0.13 | p<0.01 |
| CSF Aβ42, MMSE, Age | 0.85 ± 0.07 | 0.62 ± 0.11 | 0.77 ± 0.07 | p<0.01 |
| CSF Aβ42, MMSE, Age, Sex | 0.85 ± 0.08 | 0.62 ± 0.11 | 0.77 ± 0.07 | p<0.01 |
| **Non-invasive proposed** | | | | |
| SigLIP MRI | 0.63 ± 0.08 | 0.45 ± 0.21 | 0.63 ± 0.11 | p<0.01 |
| SigLIP MRI, CDR, Age | 0.90 ± 0.06 | 0.75 ± 0.20 | 0.83 ± 0.15 | Reference |
| SigLIP MRI , CDR, Age, Sex | 0.85 ± 0.07 | 0.52 ± 0.18 | 0.69 ± 0.10 | p<0.01 |
| SigLIP MRI, MMSE, Age | 0.89 ± 0.02 | 0.74 ± 0.05 | 0.82 ± 0.03 | p<0.01 |
| | | | | p<0.01 |

| SigLIP MRI, MMSE, Age, Sex | 0.91 ± 0.02 | 0.74 ± 0.05 | 0.82 ± 0.03 | |
|---|---|---|---|---|

**Table 3.** Classification performance of univariate and multimodal models under 1 visit. Values are calculated with a bootstrap of N = 100 repetitions and presented as mean ± standard deviation. The probability threshold was optimized with respect to balanced accuracy for every case. CSF = Cerebrospinal fluid, CDR = Clinical Dementia Rating, MMSE = Mini-Mental State Examination. P-value: Each comparison of the experimental group with the baseline reference group (SigLIP MRI, CDR, Age). p-values calculated using independent t-tests, with $p < 0.01$ indicating significant differences after correcting for multiple testing.